\documentclass{svproc}
\usepackage{url}
\usepackage{hyperref}
\hypersetup{
	colorlinks=true,
	linkcolor=red,
	filecolor=blue,      
	urlcolor=cyan,
	citecolor=blue,
}
\usepackage{algorithm}
\usepackage{algpseudocode}
\usepackage{graphicx}
\usepackage{amsmath}
\usepackage{multirow}
\usepackage{array}

\begin{document}
\mainmatter              
\title{Hulk: Graph Neural Networks for Optimizing Regionally Distributed Computing Systems}
\titlerunning{Regionally Distributed Computing Systems}  
%
\author{Zhengqing Yuan\inst{1,3} \and Huiwen Xue\inst{2} \and
Chao Zhang\inst{1} \and Yongming Liu\inst{1,4,}\thanks{Corresponding author}}

\authorrunning{Zhengqing Yuan et al.} 

\tocauthor{Zhengqing Yuan, Huiwen Xue, Chao Zhang, Yongming Liu, and Zhuanzhe Zhao}

\institute{School of Artificial Intelligence, Anhui Polytechnic University, Wuhu 241009, China,\\
\and
School of Optoelectronic Science and Engineering, Soochow University,
\\ Suzhou 215031, China
\\ \and \email{zhengqingyuan@ieee.org},\and \email{liuyongming1015@163.com}}

\maketitle              

\begin{abstract}
Large deep learning models have shown great potential for delivering exceptional results in various applications. However, the training process can be incredibly challenging due to the models' vast parameter sizes, often consisting of hundreds of billions of parameters. Common distributed training methods, such as data parallelism, tensor parallelism, and pipeline parallelism, demand significant data communication throughout the process, leading to prolonged wait times for some machines in physically distant distributed systems. To address this issue, we propose a novel solution called \textbf{Hulk}, which utilizes a modified graph neural network to optimize distributed computing systems. Hulk not only optimizes data communication efficiency between different countries or even different regions within the same city, but also provides optimal distributed deployment of models in parallel. For example, it can place certain layers on a machine in a specific region or pass specific parameters of a model to a machine in a particular location. By using Hulk in experiments, we were able to improve the time efficiency of training large deep learning models on distributed systems by more than 20\%. Our open source collection of unlabeled data:\url{https://github.com/DLYuanGod/Hulk}.
\end{abstract}
\section{Introduction}\label{intro}
In recent years, there has been a trend of scaling up deep learning models, resulting in a more robust performance in specific domains. For instance, in the field of natural language processing, large-scale text data has been used to train deep learning models such as GPT-3 (175B)~\cite{GPT3}, T5 (11B)~\cite{t5}, and Megatron-LM (8.3B)~\cite{shoeybi2019megatronlm}, which have demonstrated impressive performance. However, training these models can be quite challenging. To solve the challenges posed by large-scale deep learning models, optimization of distributed computing is crucial.

Model parallelism(MP) is a technique used to solve the problem of a model being too large to fit into the memory of a single GPU or TPU by distributing the model across multiple GPUs or TPUs. However, this approach may introduce communication challenges between GPUs or TPUs during training. On the other hand, data parallelism(DP) can improve time utilization by addressing the batch size issue during training, but it cannot resolve the problem of a model being too large for a single GPU or TPU's memory capacity.

While DP and model MP have been effective in mitigating communication volume issues in recent years, such as large minibatch SGD~\cite{goyal2017accurate}, Megatron-LM~\cite{shoeybi2019megatronlm}, Gpipe~\cite{huang2019gpipe}, and Pathway~\cite{MLSYS202298dce83d} the challenge of scheduling distributed training across machines in different regions remains unsolved. If a model like GPT-3 with hundreds of billions of parameters exceeds the memory capacity of GPUs in the current region during training, it becomes necessary to schedule machines from other regions to complete the training. This will pose several challenges:

\begin{itemize}
\item Communication latency can be very high when training is distributed across machines in different regions.
\item How can tasks be effectively allocated to different machines, such as assigning specific machines to maintain certain layers of the model's parameters (e.g., Machine 0 is responsible for Layer X) or designating machines to process specific data (e.g., Machine 2 handles Data Set Y)?
\item How can we address the issue of disaster recovery in training, such as handling scenarios where a machine fails during the process?
\item If you need to train not only a single task but also multiple tasks simultaneously, such as training both a GPT-3 and a GPT-2 model, how can you provide for these tasks?
\end{itemize}

\begin{table}[]
\caption{\label{Communication}
We measured the time it takes for our machines in three different regions to send and receive 10 words, using eight servers, and calculated the average. 
}
\begin{tabular}{lcccccccc}
\hline
\textbf{Regions} & \multicolumn{8}{c}{\textbf{Communication time to send 64 bytes (ms)}}       \\ \hline
                 & California & Tokyo & Berlin & London & New Delhi & Paris & Rome  & Brasilia \\
Beijing, China   & 89.1       & 74.3  & 250.5  & 229.8  & 341.9     & -     & 296.0 & 341.8    \\
Nanjing, China   & 97.9       & 173.8 & 213.7  & 176.7  & 236.3     & 265.1 & 741.3 & 351.3    \\
California, USA  & 1          & 118.8 & 144.8  & 132.3  & 197.0     & 133.9 & 158.6 & 158.6    \\ \hline
\end{tabular}
\end{table}

To elaborate on the first point, we collected all communication logs between the three machines and the eight servers over a three-month period. Our statistics reveal the communication time for every 64 bytes, as presented in Table \ref{Communication}. As observed in the table, the communication latency between certain nodes is high or even unfeasible. Here, the problem of communication time is difficult to solve in a distributed system without optimization.

\subsection{Contributions}
Graph data structures have been widely adopted since their introduction, as they can effectively represent interconnected structures such as social networks and knowledge graphs. Considering the tremendous success of graph neural networks~\cite{gilmer2017neural,kipf2017semi,wu2020comprehensive} in recent years, we aim to leverage this powerful capability in real-world industrial systems. With the powerful representational capability of graphs, it becomes easier to model the relevant optimization problems described in our paper. Our design choices were influenced by the types of workloads observed in actual systems. Hulk has the following features:

\paragraph{\textbf{Efficient Inter-node Communication}} Our system minimizes the impact of communication latency between machines, ensuring that each machine is assigned the appropriate task.

\paragraph{\textbf{Global Optimality}} Our model is built upon graph convolutional neural networks (GCNs)~\cite{kipf2017semi,wu2019comprehensive} to extract features from the entire graph, enabling the selection of a globally optimal solution.

\paragraph{\textbf{Disaster Recovery}} Since GCNs are utilized to assign tasks to different machines in the system, it becomes evident which tasks each machine is responsible for. Furthermore, in the event of a machine failure, the system can quickly recover the entire computation.

\paragraph{\textbf{Scalability}} If a particular machine or machines are no longer needed, you can simply remove the corresponding edge information from the graph structure.

The novelty of the proposed system lies in the utilization of graph neural networks for optimizing machine learning systems. By relying on the neural network's output values and some algorithms, the scheduling problem of the entire system can be efficiently solved.

\subsection{Engineering Challenges}
Although graph neural networks are capable of addressing tasks such as node classification~\cite{kipf2017semi,velivckovic2018graph,wang2019heterogeneous}, link prediction~\cite{zhang2018link,li2018link,schlichtkrull2018modeling}, and graph classification~\cite{kipf2017semi,ying2018hierarchical}, there is currently no suitable task that can be directly applied to our system. How to construct a suitable loss function is a crucial problem that cannot be overlooked. Regarding the representation of optimization features, such as computation time and communication time, in the graph data structure, there are also challenges that need to be addressed.

\section{Background}
This section provides a brief introduction to machine learning systems and graph neural networks.
\subsection{Machine Learning Systems}
This subsection provides a brief overview of the evolution of machine learning systems.
\subsubsection{Data Parallelism}
DP~\cite{dean2012large} is a commonly used technique in distributed training for deep neural networks, where the data is split into multiple copies and distributed to different machines for computation. Each machine calculates the loss and gradient of its assigned data and aggregates these gradients into a parameter server, which updates the model parameters. This method enables multiple machines to process large data sets in parallel, resulting in faster training speeds. 

\subsubsection{Parameter Server}
The parameter server is a distributed deep learning training method proposed by Mu Li et al.~\cite{li2014scaling} that addresses the communication bottleneck problem in training large-scale deep learning models. It achieves this by placing the gradient aggregation and parameter updating process on the server side, and the computational nodes only need to send the locally computed gradient information to the server. This approach reduces communication overhead and improves training efficiency.

\subsubsection{Megatron-LM}
Megatron-LM~\cite{shoeybi2019megatronlm} combines model parallelism and data parallelism by dividing the model parameters into multiple parts, each trained on a different GPU. This allows for larger models to be used as each GPU only needs to focus on computing a part of the model using model parallelism. Data parallelism is used to assign different batches to different GPUs for processing, which improves training efficiency.

The training objective of Megatron-LM is to minimize the negative log-likelihood of the target sequence given the input sequence, which is expressed as:
\begin{eqnarray*}
L(\theta) = - \sum_{t=1}^{T} \log P(y_t | y_{<t}, x; \theta)
\end{eqnarray*}
where $T$ is the length of the sequence, $y_t$ is the target token at time step $t$, $y_{<t}$ are the tokens before time step $t$, $x$ is the input sequence, and $\theta$ represents the model parameters.

\subsubsection{Gpipe}
In Gpipe~\cite{huang2019gpipe}, the model is split into sub-models, each assigned to a different GPU. DP concatenates Micro-batches along the pipeline to pass data and gradients between GPUs, enabling pipeline parallelism~\cite{dally1996pipeline}. The training process in Gpipe can be expressed as the following equation:

\begin{eqnarray*}
\Delta W_{i,j}=\eta\sum_{k=1}^{K}(\nabla_{W_{i,j}}L(f^{i,j}(x_{k}^{i,j}),y_{k}^{i,j})+\sum_{l=j+1}^{M}\nabla_{W_{i,l}}L(f^{i,l}(x_{k}^{i,l}),y_{k}^{i,l}))
\end{eqnarray*}
where $W_{i,j}$ denotes the weight parameter of the $j$th layer of the $i$th submodel, $\Delta W_{i,j}$ denotes the corresponding parameter update, $\eta$ denotes the learning rate, $K$ denotes the number of Micro-batches, $f^{i,j}$ denotes the forward propagation function of the $j$th layer of the $i$th submodel, $x_ {k}^{i,j}$ denotes the $k$th Micro-batch of the $j$th layer in the $i$th sub-model, $y_{k}^{i,j}$ denotes the label of the $k$th Micro-batch.

\subsection{Graph Neural Networks}
Graph Neural Networks (GNNs)~\cite{scarselli2008graph,zhou2018graph,zhang2018scalable,bui2021structural,hamilton2017inductive} are a type of neural network designed to work on graph-structured data, where nodes represent entities and edges represent relationships between them. They have become popular in recent years due to their ability to capture complex relationships and patterns in data, making them useful for tasks such as node classification, link prediction, and graph classification.

\subsection{Graph Convolutional Networks}
 Graph Convolutional Networks (GCNs)~\cite{kipf2017semi} are a type of deep learning model designed to work on graph-structured data. They use convolutional operations to aggregate information from neighboring nodes and update node representations. The key formulas for GCNs include the graph convolution operation, which calculates the node representation updates, and the graph pooling operation, which aggregates information across multiple nodes.

\begin{equation}
\mathbf{v}^{(l+1)} = \sigma\left(\sum_{u \in \mathcal{N}(v)}\frac{1}{c_{u,v}} W^{(l)}\mathbf{u}^{(l)}\right)
\label{eq:one}
\end{equation}
where $\mathbf{v}^{(l)}$ represents the feature representation of node $v$ at layer $l$, $\mathcal{N}(v)$ denotes the set of neighbors of node $v$, $W^{(l)}$ is the weight matrix at layer $l$, $\sigma$ is the activation function, and $c_{u,v}$ is a normalization factor that depends on the number of neighbors of node $u$ and $v$. This formula is used to iteratively compute the feature representations of nodes in a graph using neighborhood information.

\section{Data Representation}
To better address the issues raised in Section~\ref{intro}, it is important to select an appropriate data structure to represent the system parameters.We adopt a graph-based data structure to represent our system parameters, with each node (denoted as $v$) representing a machine in a different region. Each node has unique features that include its geographic location, computational capacity, and GPU memory. The edges (denoted as $e$) between nodes denote the possibility of communication between the two connected machines, with the weight of each edge representing the time in milliseconds required to transmit each 64-byte message.

\begin{figure}[h]
\centering
\includegraphics[scale=0.58]{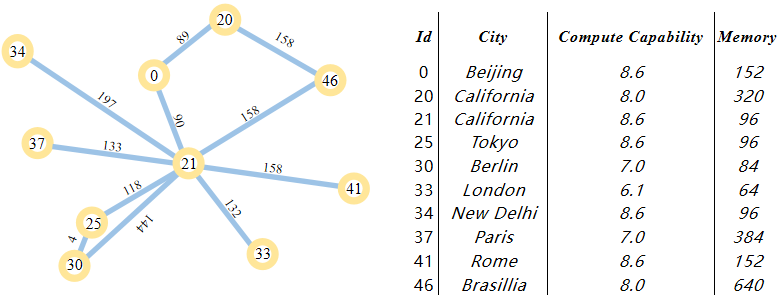}
\caption{In this figure, the graph topology is visualized on the left, while the characteristics of each node are indicated on the right. Where computing power is determined based on Nvidia's official website\protect\footnotemark, and memory refers to the total memory across all GPUs on each machine.}\label{Data}
\end{figure}

As depicted in Figure \ref{Data}, we randomly selected eight machines to construct a graph, where the edge weight represents the communication time, and the node features are embedded in the corresponding vector space.

\footnotetext{\url{https://developer.nvidia.com/cuda-gpus}}
For example, node 0 can be represented as $v_{0}=\left \{'Beijing',8.6,152\right \} $. Then we embed the node information using the following formula:
\begin{equation}
\mathbf{v}^{(0)}=\mathbf{x}_{v}
\label{eq:two}
\end{equation}
where $\mathbf{v}^{(0)}$ denotes the initial feature vector of node $v$ and $\mathbf{x}_{v}$ denotes the input feature vector of node $v$.

The node-to-node edges we represent by the adjacency matrix. The weight of an edge in the adjacency matrix is equal to the communication time between two corresponding nodes. The values for unconnected edges are set to 0, and the diagonal values in this matrix are all 0. Similarly, we then perform the edge information embedding with the following equation:

\begin{equation}
e_{vu} = g\left(\mathbf{e}_{vu}, \mathbf{u}, \mathbf{v}, \mathbf{\Theta}_e\right)
\label{eq:two}
\end{equation}
where $e_{vu}$ denotes the edge feature between node $v$ and node $u$, $\mathbf{e}_{vu}$ is the feature vector of edge $vu$, $\mathbf{u}$ and $\mathbf{v}$ are the feature vectors of node $u$ and node $v$, respectively, $g$ is a learnable function and $\mathbf{\Theta}_e$ is its argument. We then sparsely label this subgraph to enable the neural network to learn the contents of the graph in a supervised manner.

\section{Methods}\label{Methods}
The typical tasks of graph neural networks, such as node classification, do not utilize edge information and only leverage the graph topology. In real-world cases, the information carried by edges is often crucial, such as edge weights and directed edges. To incorporate edge information into nodes, we aim to perform edge pooling, which involves aggregating or pooling edges of neighboring nodes at each node to create a unified node representation that contains edge information. This is expressed in the following equation:

\begin{equation}
\mathbf{v}^{(l+1)} = \sigma\left(\sum_{u \in \mathcal{N}(v)} f(\mathbf{v}^{(l)}, \mathbf{u}^{(l)}, e_{vu})\right)
\label{eq:two}
\end{equation}

Where $\mathbf{v}^{(l+1)}$ represents the feature vector of node $v$ in layer $l+1$, $\sigma$ is the activation function, $\mathcal{N}(v)$ denotes the set of neighboring nodes of node $v$, $\mathbf{u}^{(l)}$ represents the feature vector of node $u$ in layer $l$, and $f$ is a learnable function used to merge features of nodes and edges into new features of node $v$.

\begin{figure}[]
\centering
\includegraphics[scale=0.35]{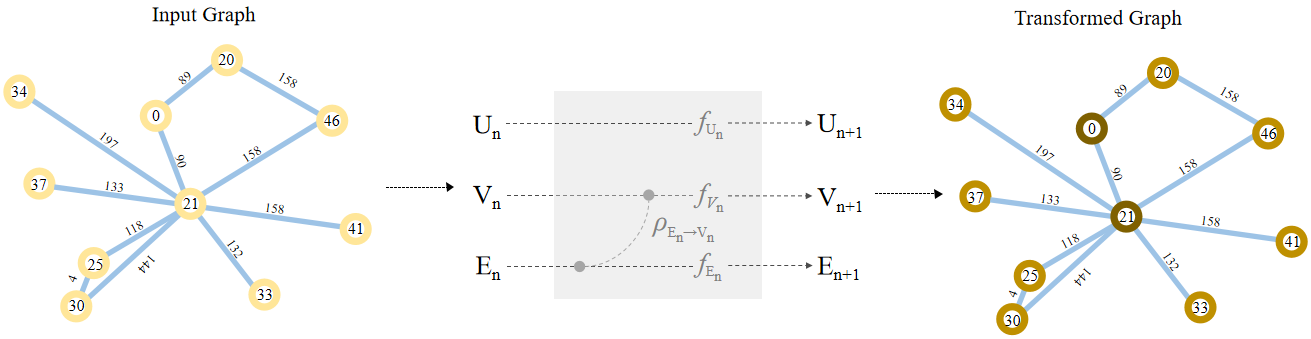}
\caption{The edge pooling operation of the above figure~\ref{Data}. where $U$ represents the information of the whole graph and $f$ is the respective linear layer.}\label{Trans}
\end{figure}

As depicted in Figure \ref{Trans}, this is the first layer of the constructed network structure($l=0$) that enables nodes to encode edge information.

\begin{figure}[h]
\centering
\includegraphics[scale=0.40]{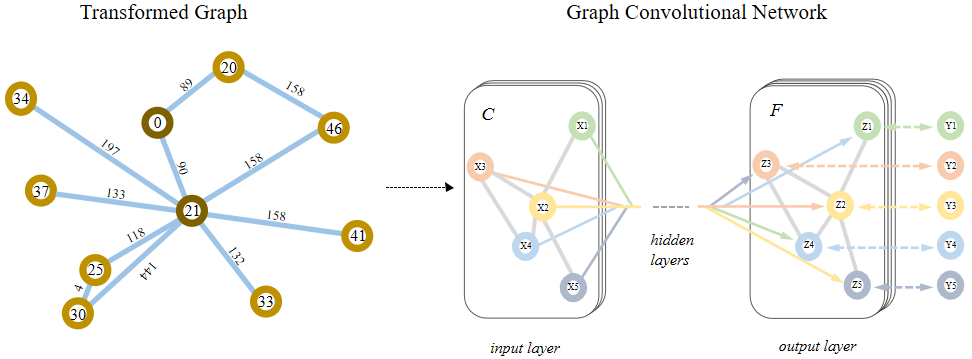}
\caption{The transformed graph data are entered into GCNs for forward propagation. }\label{GCNs}
\end{figure}

After the edge features are embedded into node features, we can use the resulting transformed graph as input for a standard node classification task and train it using a graph convolutional neural network or graph attention network. As shown in Equation \ref{eq:one}. If we want to build N-layer GCNs with our $l=2,3,4\cdots N+1$.

As shown in Figure \ref{GCNs}, Y represents the category of the classification, i.e., what tasks are appropriate.

Then we calculate its loss using the cross-entropy loss function~\cite{goodfellow2016deep}:

\begin{equation}
\mathcal{L} = -\sum_{i=1}^{|\mathcal{Y}|} Y_{i}\log \hat{Y}_{i}
\label{eq:three}
\end{equation}
Here, $\mathcal{Y}$ denotes the set of all labels, $Y_{i}$ denotes the true label of node $i$, and $\hat{Y}_{i}$ denotes the predicted label of node $i$. Then back propagation is performed to update the network parameters.

\begin{figure}[h]
\centering
\includegraphics[scale=0.50]{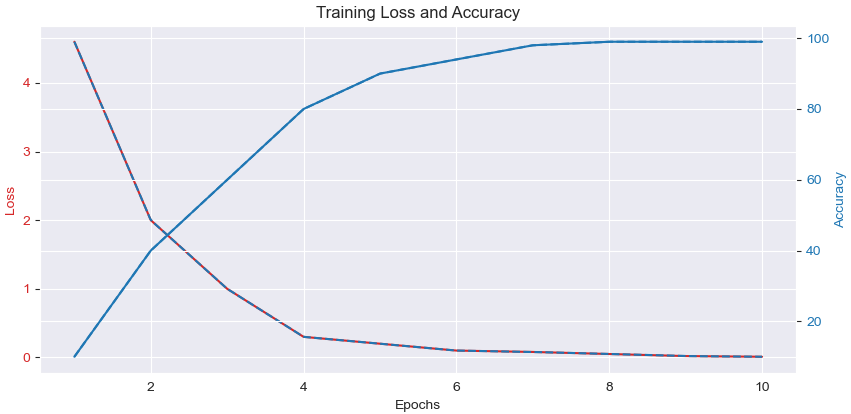}
\caption{Loss rate and accuracy line charts for 10 steps of training on this data. The parameters of GCNs are 188k and the learning rate is 0.01. }\label{Train}
\end{figure}
As depicted in Figure \ref{Train}, we observed that the accuracy peaked at 99\% during the sixth training step. 

\section{Structure}
In this section, we build our system based on the GCNs trained in the previous section \ref{Methods} and solve the problem presented in section~\ref{intro}.

\subsection{Efficiency}
We now have two tasks to perform. The first involves training the BERT-large model~\cite{devlin2019bert}, while the second involves training the GPT-2 model~\cite{radford2019language}. As the largest GPT-2 model (1.5B parameters) is significantly larger than BERT-large (340M parameters), it is important to carefully allocate tasks to each machine in a sensible manner. The ratio of the number of parameters in GPT-2's largest model (1.5B) to BERT-large (340M) is approximately 4.4:1. Based on this information, we instruct the graph neural network to classify the classes according to this scale and optimize the communication time within each class. Also, we need to consider the memory and computing power characteristics of each machine. 

\begin{algorithm}[H]
\caption{Task Assignments}
\label{alg1}
\begin{algorithmic}[1]
\Require Graph Data $G_{1}$, Trained Graph Neural Network $F$, Number of Tasks $N$, Minimum Memory Threshold $M_{n}$ for Each Task
\Ensure Task Assignments for Each Graph Data
\State $C \gets 0$
\If{$G_{1}$ does not meet the requirements of all tasks}
\State Jump out of the algorithm and report an error.
\EndIf
\For {i in range(1, N)}
\State $G_{i}, G_{i+1} \gets F(G_{i})$
\State Assign the smaller graph $G_{i}$ to a task with the appropriate minimum memory threshold $M_{n}$
\If{$G_{i}$ does not meet the requirements of the all task}
\State $C \gets i$ and Continue
\If{$C>=1$}
\State $G_{i} \gets G_{i} + G_{C}$ 
\State Assign the smaller graph $G_{i}$ to a task with the appropriate minimum memory threshold $M_{n}$
\State $C \gets 0$
\EndIf
\EndIf
\If{$G_{i+1}$ does not meet the requirements of the all task}
\State Break and Provide a prompt and wait for other tasks to complete before proceeding with training.
\EndIf
\EndFor
\end{algorithmic}
\end{algorithm}

We use Algorithm \ref{alg1} to schedule multiple tasks, but it can also be used to determine superiority if there is only one task. Based on the computational power, memory and communication efficiency features, as well as the integration into node information, we only need to determine whether it is appropriate.

\begin{figure}[h]
\centering
\includegraphics[scale=0.50]{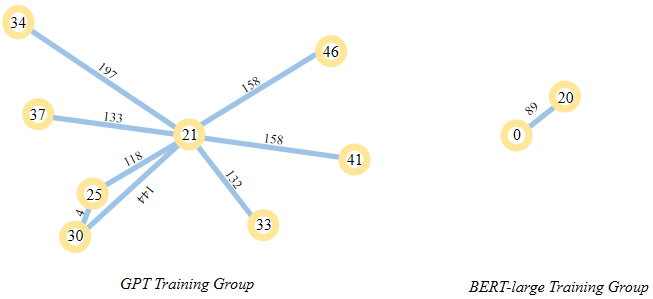}
\caption{The data in Figure \ref{Data} are grouped using Algorithm \ref{alg1}. The left panel is the training group of GPT-2 and the right panel is the Bert-large training group. }\label{Group}
\end{figure}

Figure \ref{Group} demonstrates that the basic graph neural network is capable of carrying out classification tasks effectively and emulating human thought processes.

\subsection{Scalability}
If we need to add one or more machines to this system, we can simply define their $\left \{City, Compute Capability, Memory\right \}$ and connect them to the existing nodes that can communicate with them using weights.

\begin{figure}[h]
\centering
\includegraphics[scale=0.50]{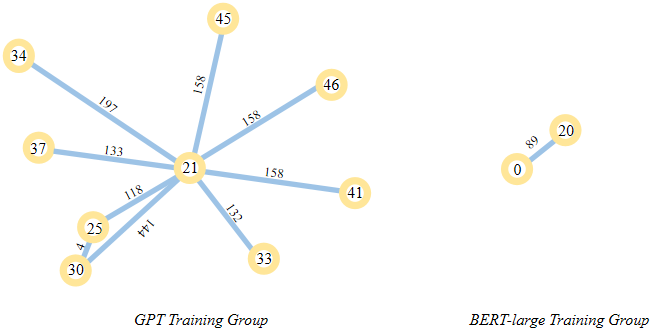}
\caption{Join the machine with id 45 and make assignments. }\label{Add}
\end{figure}

As shown in Figure \ref{Add}, the machine with id 45$\left \{Rome, 7, 384\right \}$ in the dataset was added to the Hulk system and still works fine.

\section{Experimentation and Evaluation}
In this section, we test the Hulk system using multiple deep learning tasks in real industries with 46 high-performance GPU servers.

\subsection{Experimental Setting}
We have a total of 46 servers distributed across different countries and regions, with a combined total of 368 GPUs of various models such as NVIDIA A100, NVIDIA A40, NVIDIA V100, RTX A5000, GeForce GTX 1080Ti, GeForce RTX 3090, and NVIDIA TITAN Xp. And, we calculated the average of 10 communications between these machines over a 3-month period. Due to network policy restrictions in different countries, there are certain machines that are unable to communicate with each other. We adopt the parameter settings provided in the original paper for the training process.

\subsection{Data Building}
\begin{figure}[h]
\centering
\includegraphics[scale=0.30]{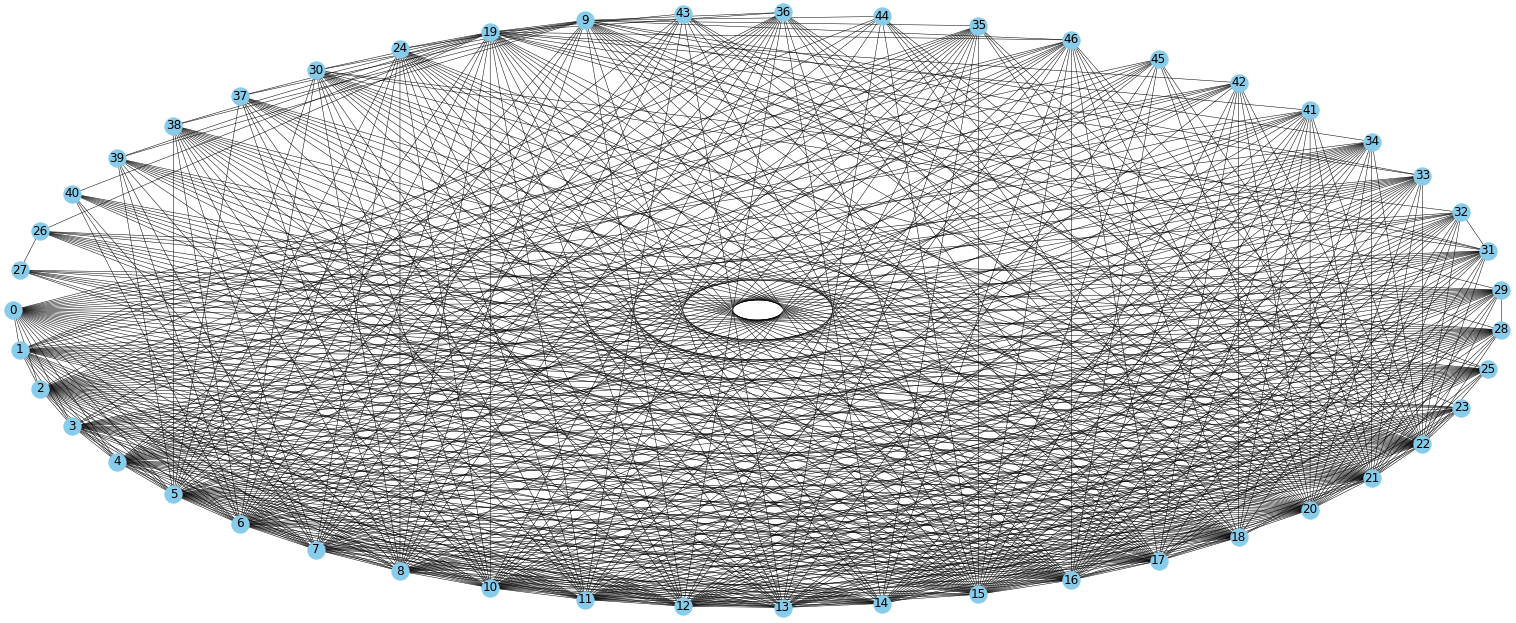}
\caption{46 servers construct the graph structure data. }\label{dA}
\end{figure}

We use networkx~\cite{hagberg2008exploring} library to build our graph structure data and visualize it as shown in Figure \ref{dA}. Additionally, we need to read the adjacency matrix of this data and consider the corresponding feature embedding representation.

\subsection{Task Assignment}
The four tasks we aim to train in this system are OPT (175B)~\cite{iyer2023optiml}, T5 (11B), GPT-2 (1.5B), and BERT-large (350M). 

We need to classify all nodes into four distinct classes based on their characteristics and then deploy distributed algorithms tailored to each class.

\begin{table}[h!]
\centering
\caption{Model Node Allocation}
\label{tab:model_node_allocation}
\begin{tabular}{cc}
\hline
\textbf{Model} & \textbf{Nodes} \\
\hline
OPT (175B) & 0, 1, 2, 3, 4, 20, 21, 22, 23, 24, 27, 28, 29, 30, 31 \\
T5 & 5, 6, 7, 8, 9, 10, 11, 12, 13, 14 \\
GPT-2 & 15, 16, 17, 18, 19, 25, 26, 32, 33, 34 \\

BERT-large & 35, 36, 37, 38 \\
\hline
\end{tabular}
\end{table}

As presented in Table \ref{tab:model_node_allocation}, we feed the graph data into the graph neural network, which was trained in Section \ref{Methods} and employs Algorithm \ref{alg1}, to derive node classification information. To handle the nodes in each class with different computational performance and memory, we utilize Gpipe to train the model in parallel. Depending on the computational power and memory of each node, we determine which part of the model it will handle.

\subsection{Evaluation}
To validate the performance of the Hulk system, we have chosen three commonly used distributed computing algorithms for evaluation.

\paragraph{\textbf{System A}} It utilizes all available machines for training while discarding any machine that does not have sufficient memory to accommodate the entire model. It utilizes data parallelism to distribute the batch size across multiple machines, thereby enabling simultaneous training of the model on each machine.

\paragraph{\textbf{System B}} It utilizes Gpipe for parallelism, assigning a certain layer of the model to a particular machine until the entire model is distributed across all machines.

\paragraph{\textbf{System C}} It employs tensor parallelism with Megatron-LM across the entire system, requiring all machines to be utilized for model training.

\begin{figure}[h]
\centering
\includegraphics[scale=0.47]{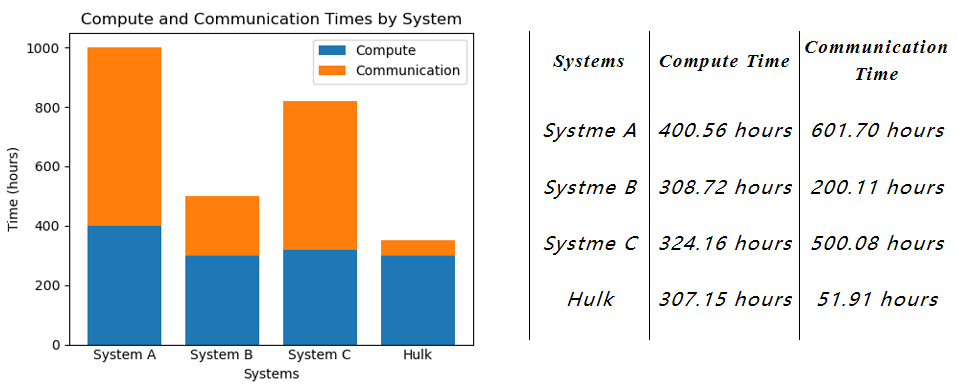}
\caption{Communication time and calculation time on four models of the four the 4 systems. }\label{Result}
\end{figure}

\paragraph{Result} As shown in Figure \ref{Result}, the Hulk system can greatly reduce communication time and thus the overall training time. This illustrates that Hulk is effective in dividing the nodes into a specific model for training.

\begin{figure}[h]
\centering
\includegraphics[scale=0.60]{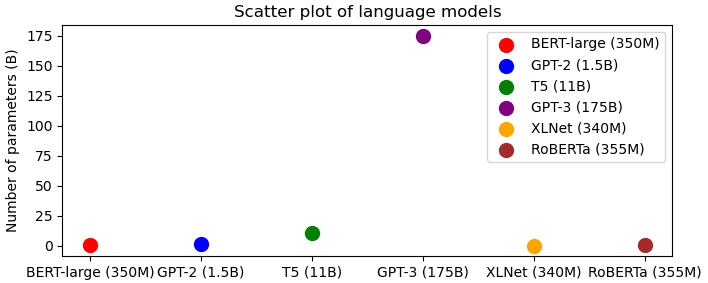}
\caption{Language Model Parameters}\label{model}
\end{figure}

If we need to train 6 models, the parameters of each model are shown in Figure \ref{model}. Among them, the parameters of RoBERTa~\cite{robust} are 355M and the parameters of XLNet~\cite{Xlnet} are 340M.

\begin{figure}[]
\centering
\includegraphics[scale=0.47]{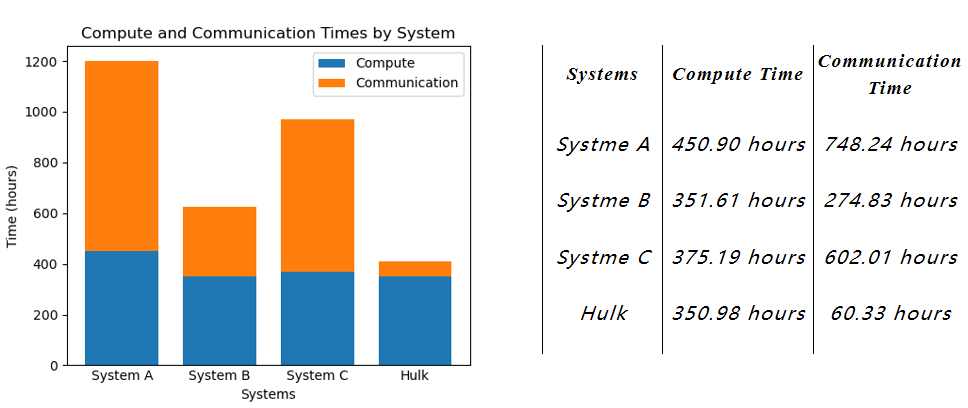}
\caption{Communication time and calculation time on six models of the four systems. }\label{Result2}
\end{figure}

\paragraph{Result} As illustrated in Figure \ref{Result2}, when the system needs to handle multiple tasks, the gap in communication time becomes more apparent. Our Hulk system is able to effectively reduce communication time (Because the GPT-3 (175B) model is not open source, we use the OPT (175B) with equivalent parameters instead).

\section{Conclusion}
In this article, we introduce our novel solution, Hulk, which optimizes regionally distributed computer systems by tackling the challenges of scheduling distributed training across machines in different regions. Our real-world industrial solution, Hulk, utilizes graph neural networks with powerful representation capabilities to enhance communication efficiency between GPUs or TPUs across different countries or regions during training. With its efficient communication, global availability, fast recovery, and excellent scalability, Hulk stands out as a powerful tool for optimizing regionally distributed computer systems. The results demonstrate a significant increase in the efficiency of distributed training, crucial for the success of large-scale deep learning models. Overall, the use of Hulk can streamline the model deployment process and benefit researchers and practitioners seeking to optimize communication efficiency.

\section*{Acknowledgement}
The authors gratefully acknowledge the support of the AIMTEEL 202201 Open Fund for Intelligent Mining Technology and Equipment Engineering Laboratory in Anhui Province and the Anhui Provincial Department of Education Scientific Research Key Project (Grant No. 2022AH050995). The financial assistance provided by these projects was instrumental in carrying out the research presented in this paper. We would like to thank all the members of the laboratory for their valuable support and assistance. Without their help, this research would not have been possible. Finally, we would like to express our gratitude to the Anhui Polytechnic University for providing the necessary facilities and resources for this study.

%
%

%
\bibliography{custom}  
\bibliographystyle{bibtex/spmpsci}  

\end{document}